\documentclass[12pt]{iopart}

\usepackage{iopams}
\usepackage{graphicx}

\newcommand{\diff}{\mathrm{d}}

\begin{document}

\title{The quantum-to-classical transition: contraction of associative products}

\author{A. Ibort$^{a,f}$, V.I. Man'ko$^{b,c}$, G. Marmo$^{d,e}$, A. Simoni$^{d,e}$,
C. Stornaiolo$^e$, F. Ventriglia$^{d,e}$}

\address{$^a$Departamento de Matem\'{a}ticas, Universidad Carlos III de Madrid,
Avda. de la Universidad 30, 28911 Legan\'{e}s, Madrid, Spain.}

\address{$^b$Lebedev Physical Institute, Leninskii Prospect 53, Moscow 119991, Russia.}

\address{$^c$Moscow Institute of Physics and Technology (State University), Dolgoprudny, Moscow Region, Russia.}

\address{$^d$Dipartimento di Fisica ``E. Pancini" dell' Universit\`{a} ``Federico II" di Napoli, Complesso Universitario di Monte S. Angelo, via Cintia, 80126 Naples, Italy.}

\address{$^e$Sezione INFN di Napoli, Complesso Universitario di Monte S. Angelo, via Cintia, 80126 Naples, Italy.}

\address{$^f$Instituto de Ciencias Matem\'{a}ticas (CSIC - UAM - UC3M - UCM), Nicol\'{a}s Cabrera,13-15, Campus de Cantoblanco, UAM, 28049, Madrid, Spain.}

\ead{albertoi@math.uc3m.es, manko@sci.lebedev.ru, marmo@na.infn.it, simoni@na.infn.it, cosmo@na.infn.it, ventriglia@na.infn.it}

\begin{abstract}  
The quantum-to-classical transition is considered from the point of view of
contractions of associative algebras.    Various methods and ideas to deal with contractions of associative algebras are discussed that account for a large family of examples.   As an instance of them, the commutative algebra of functions in phase space, corresponding to classical physical observables, is obtained as a contraction of the Moyal star-product which characterizes the quantum case.
Contractions of associative algebras associated to Lie algebras are discussed, 
in particular the Weyl-Heisenberg and  $SU(2)$ groups are considered.
\end{abstract} 

\pacs{03.65-w, 03.65.Wj} 
\vspace{2pc}
\noindent{\it Keywords}: Quantum-to-classical transition, decoherence dissipation, Associative algebras, contraction.
\submitto{\PS}

\maketitle


\section{Introduction}

In this paper ``decoherence'' dissipation, or transition to classicality, will be
discussed within the framework of contraction of algebraic structures by exploring a contraction procedure of associative algebras (see for instance \cite{CGRM,CGM} and references therein).

The quantum-to-classical transition is a fascinating subject both from the physical, mathematical and philosophical point of view.   At the dawn of the quantum theory, it entered the picture as the ``correspondence principle" \cite{bohr1,Bimonte:2002jp} and the so-called wave-particle duality was elevated to the ``principle of complementarity'' \cite{bohr2} but, more formally, in the Schr\"odinger picture, the quantum-to-classical transition can be described as the transition from the  Schr\"odinger to the Hamilton-Jacobi equations of classical mechanics for a given state represented as a wave function (see \cite{Esposito:2002wh}).

\begin{figure}[h]
  \centering
   \includegraphics[width=14cm]{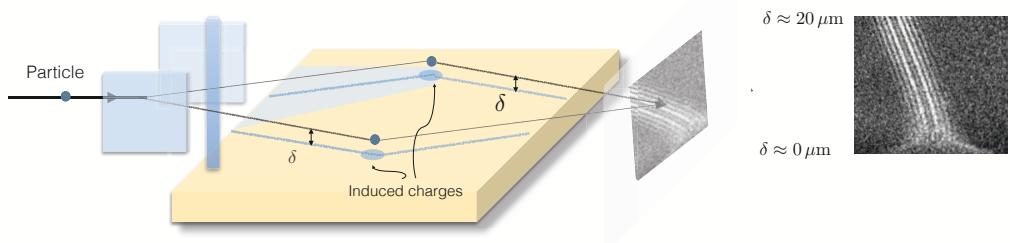}
  \caption{Realization of decoherence experiment, after Sonnetag and Hasselbatch \cite{sonnentag1} (auxiliary coils and equipment has been omitted from the picture).}\label{fistra}
\end{figure}

Recently, the double-slit interference experiment with electrons  \cite{merli1,merli2}, has been enhanced by using a plate of conductor material with very high resistivity \cite{sonnentag,sonnentag1} (Figure \ref{fistra}).  By changing the distance $\delta$ of the conductor from the path of the electrons it is possible to decide which way the electron went by exploiting the Joule effect due to the motion of the mirror charge.  Here the visibility parameter gives a measure of the ``classicality" of the state and can also be considered as a ``deformation'' parameter.

In the same vein and in the Heisenberg picture, the quantum-to-classical transition can be described in terms of the quantizer-dequantizer formalism as a $\hbar \to 0$ limit process in the algebra structure $f_A \ast_\hbar f_B$ were $f_A$ is the phase space function associated to the quantum observable $A$  and the product is the $\hbar$ depending star-product $f_A \ast_\hbar f_B = f_{AB}$ which turns out to be non-local and non-commutative  (see for instance \cite{OlgaJPA}). 

Then, the non-locality could be ``measured" by considering the extension of the region where  the expectation values of the selected observables are sensibly different from zero.  In this setting the transition from a non-local product to a local one is well described by the Moyal approach in terms of higher order bi-differential operators which, at zeroth order in the Planck's constant, fixes the pointwise product and, at first order in $ \hbar$, gives the Poisson bracket (see for instance \cite{moyal49} \cite{Gr} \cite{fate} and references therein).

In this work we will concentrate mostly on the non-commutativity of the family of selected measurable operators, considering that the commutative case corresponds to the classical situation and the less commutative, `less classical' or `more quantum' the system, and the states describing it, is.
Thus, in a sense, we take the point of view that ``full classicality" is achieved when all observables commute pairwise. 

In the intermediate stages, a greater degree of ``classicality" may be ascribed to a larger number of observables which pairwise commute.  Thus for instance, in the previously described double-slit experiment with a conductive plate, we may consider that the commutation relation of the position observables $\hat{x}$, $\hat{y}$ for the electron will have the form $[\hat{x}, \hat{y}] \sim \delta$ where $\delta$ is the distance from the plane of the electron to the plate.   Hence the limit $\delta \to 0$ will correspond to the classical situation (no interference) while as $\delta$ increases, the quantum nature of the system prevails.  Notice, however, that this is not the standard description of the double-slit experiment where the quantum position observables, $\hat{x}$ and $\hat{y}$, do commute.

The mathematical construction which takes us from a maximally non-commuting to a larger and larger family of pairwise commuting observables goes under the name of `contraction procedure' (see for instance,  \cite{CGRM,Chruscinski:2011yv}).   Therefore we will consider here the contraction procedure applied to associative structures to describe the quantum-to-classical transition, and in the case of the associative $\ast_\hbar$ product above, it will be applied to its associative kernel, which appear when operators are replaced by functions on some measure space and the operator product is replaced by a star-product as in the quantizer-dequantizer formalism (see below, Sects. \ref{sec:quantizer}, \ref{sec:moyal}). 

Before entering the details of this idea, let us first briefly recall how contraction procedures entered Physics by the hands of In\"{o}n\"{u} and Wigner \cite{inonu}, but also Segal \cite{segal} and Saletan \cite{saletan}, in the early Fifties.

Starting from the consideration that Minkowski space-time `reduces' to Gali\-leo's
space-time when the speed of light tends to infinite,  in 1953 In\"{o}n\"{u} and Wigner
suggested that the inhomogeneous Lorentz group should give as a limiting
case, when the velocity of the light goes to infinity, the Galilei group. They introduced what nowadays is called the contraction procedure for Lie algebras.

Thus Wigner-In\"{o}n\"{u}'s procedure \cite{inonu} to contract the  $ SU(2) $ algebra to get the motion group $ E(2) $ algebra is based on the use of a parameter depending linear transform $T_\epsilon$ in the vector space of the Lie algebra of the group.  It is assumed that for any value $ \epsilon\neq 0 $ the linear transformation is invertible and becomes singular for $ \epsilon=0 $. If the appropriate limit on the structure constants exists, we get new structure constants that, provided they satisfy Jacobi's identity, could correspond eventually to a different Lie algebra structure.  

Of course, similar considerations can be made with respect to the limit of Planck's constant $\hbar$ 
going to zero when considered within the structure constants defining the Lie algebra of the Heisenberg-Weyl group.   In this case the Heisenberg-Weyl algebra contracts to the corresponding Abelian algebra.  Even more generally, if we have a family of generators closing on a finite dimensional Lie algebra, then
Ado's theorem provides the possibility to realize this Lie algebra by means of an algebra of matrices, hence to
extend the Lie algebra structure to an associative algebra of matrices.  
Thus it would be possible to study contractions  of algebras of matrices using Ado's theorem under this guise.

Here we apply the same procedure when the structure constants are those of an associative algebra. In particular the associative algebra may be a group algebra or a groupoid algebra. More generally it could be a star-product algebra originating from some operator algebra via a quantizer-dequantizer procedure \cite{OlgaJPA} \cite{Ibort} \cite{Ibort:2013jaa} \cite{AssociaMoyal}.   Thus in the quantizer-dequantizer scheme it seems quite natural to consider the contraction procedure at the level of kernel functions that play the role of the structure constants of the associative product, clearly extending in this manner the Lie
algebra contraction procedure considered by In\"{o}n\"{u} and Wigner. 

Similar problems have been already considered in the tomographic picture \cite{Ibort} in
connection, for instance, with the fate of the complex structure in going from quantum to
classical mechanics \cite{fate}. 

Here we would like to take up first the problem
of contractions at the level of Weyl systems considered as unitary
representations of the Heisenberg-Weyl group, in particular in connection
with the group algebra associated to a given Lie group and the convolution
product.  Let us mention here that related papers, connected with dissipation or decoherence seen as a
contraction procedure, have appeared recently \cite{Chruscinski:2011yv,Chrubis}.

Summarizing, we may say that in this short paper we will try only to outline the main ideas and difficulties which arise in trying to implement the ``contraction procedure" on
associative algebras, a more systematic analysis should be postponed to a
future paper.

The paper is organized as follows. In Sect. \ref{sec:weyl}, and as a motivation example, we present the quantum-to-classical transition as a contraction in the Gr\"{o}newald kernel of the star-product of functions on phase-space associated with the Heisenberg-Weyl group algebra  when $\hbar\to 0$ \cite{Gr}.  Moreover a general scheme to find new structure constants in the framework of the quantizer-dequantizer formalism will be analized too. In Sect. \ref{sec:alternative} we will discuss a general and fairly simple description for the contraction procedure for associative algebras and we will illustrate it in some simple examples.  In Sect. \ref{sec:conclussion}, finally, some perspectives and conclusions are drawn.


\section{The case of infinite dimensional algebras:  Weyl systems, the quantizer- dequantizer formalism and the contraction of the Moyal product}\label{sec:weyl}


\subsection{Weyl systems}

We discuss in this section the properties of Weyl systems \cite{Weyl27} \cite{Wigner32} and Weyl quantization \cite{Berezin:1966nc} in the context of the star-product formalism provided by a class of quantizer-dequantizer operators \cite{OlgaJPA}. The Weyl quantization is a particular case of the star-product formalism to be discussed later on (see Sect. \ref{sec:quantizer}), but the material of this section can be read independently of it.  Planck's constant  $\hbar$ will be set to 1 in the formulae but will be reintroduced when necessary for considering the limit $ \hbar\to 0 $.

A Weyl system $W$ is defined as a map from an Abelian complex vector group $E$  to the group of unitary operators on some Hilbert space $\mathcal{H}$ (we restrict for simplicity to one-dimensional complex spaces) such as:
\begin{equation}
\widehat{W}(z) =\exp \left[ z  \hat{a}^{\dagger} - z^{\ast}  \hat{a} \right]
\end{equation}
where 
 \begin{equation}
\hat{a}= \frac{\hat{q}+ i \hat{p}}{\sqrt{2}}, \quad \hat{a}^{\dagger}= \frac{\hat{q}- i \hat{p}}{\sqrt{2}}, \quad \left[ \hat{q},\hat{p} \right]=i,
\end{equation}
are boson creation and annihilation operators with commutation relation: 
\begin{equation}
[\hat{a},\hat{a}^{\dag}]=1\, ,
\end{equation}
and  $ \hat{q} $ and $ \hat{p} $ are dimensionless position and momentum operators, respectively. The complex coordinate:
\begin{equation}
z=\frac{1}{\sqrt{2}}\left(q+ip \right), 
\end{equation}
where the pair $ (q,p) $ determines a point in the complex one-dimensional Abelian vector group $E$ determines its complex structure. 

Weyl quantization provides a bijective map from functions on $E$ into operators acting on the Hilbert space $\mathcal{H}$, i.e., it maps phase-space functions $f = f(q,p)$ into operators $\hat{f}$ on $\mathcal{H}$ (e.g., the quantum harmonic oscillator states define the space of square integrable functions on the line), by means of the formula:
  \begin{equation}\label{w5}
  \hat{f}=\frac{1}{\pi} \int f(q,p)\widehat{W}(2z)\widehat{\mathcal{P}} \,\, \mathrm{d}q \mathrm{d}p.
  \end{equation}
  Here $\widehat{\mathcal{P}}$ is the parity operator, i.e., $\widehat{\mathcal{P}}\psi(x)=\psi(-x) $.
  
  The inverse transform provides a map from operators $ \hat{f} $ to functions $ f(q,p) $, given by: 
  \begin{equation}\label{w6}
  f(q,p)=2 \,  \mathrm{Tr} \, (\hat{f}\, \widehat{W}(2z)\widehat{\mathcal{P}}) \, .
  \end{equation}
 
 Weyl's quantization provides an example of a realization of a star-product scheme in which the functions on  phase-space are multiplied according to Moyal's  deformed product \cite{moyal49} \cite{Berezin:1966nc}, i.e., a non-local and non-commutative product of the form:
  \begin{equation}\label{starproduct}
  \left(f_{1}\ast f_{2} \right) (q,p)= \int f_{1}(q_{1},p_{1}) f_{2}(q_{2},p_{2})K(q_{1},p_{1},q_{2},p_{2},q,p) \, \, \mathrm{d}q_{1}\mathrm{d}p_{1}\mathrm{d}q_{2}\mathrm{d}p_{2} \, .
  \end{equation}  
 Moyal's product can also be described by using a pair of operators, the dequantizer $\widehat{U}(q,p)$ and the quantizer  $\widehat{D}(q,p)$ operators given by: 
   \begin{equation}
   \widehat{U}(q,p) =2\, \widehat{W}(2z) \widehat{\mathcal{P}}=2\, \widehat{W}(\sqrt{2}q+i\sqrt{2}p) \widehat{\mathcal{P}}, \qquad  \widehat{D}(q,p)=\frac{1}{2 \pi} \widehat{U}(q,p) 
   \end{equation}
  that satisfy the biorthogonality relation: 
  \begin{equation}\label{w8}
 \mathrm{Tr} (\widehat{U}(q,p)\, \widehat{D}(q',p') ) =\delta(q-q')\delta(p-p')  \, .
  \end{equation}
  
  Hence formulae (\ref{w5}) and (\ref{w6}) can be rewritten in terms of the previous quantizer and dequantizer operators as:
  \begin{equation}\label{fhatweyl}
  \hat{f}=\int \diff q \diff p \, \widehat{D}(q,p) f(q,p)
  \end{equation}
  and
  \begin{equation}\label{fweyl}
  f(q,p)=\mathrm{Tr} (\hat{f}\,\widehat{U}(q,p) )
  \end{equation}
  The kernel $K$ of Moyal's star-product in Eq. (\ref{starproduct}), can be written again in terms of the quantizer and  dequantizer  operators as:
  \begin{equation}\label{kernelweyl}
  K(q_{1},p_{1},q_{2},p_{2},q_{3},p_{3})= \mathrm{Tr} (\widehat{D}(q_{1},p_{1})\widehat{D}(q_{2},p_{2})\widehat{U}(q_{3},p_{3})) \, .
  \end{equation}


\subsection{Quantizer-dequantizer formalism}\label{sec:quantizer}

The quantizer-dequantizer formalism used in the previous section to express Weyl's quantization
in a compact form, Eqs. (\ref{fhatweyl})-(\ref{kernelweyl}), can be nicely expressed in a 
general setting as follows.

Let $\widehat{U}(x),\widehat{D}(x)$ be two maps, called the dequantizer and quantizer respectively, from a measure space $X$ in the space of operators describing a quantum system and such that they satisfy the biorthogonality relation:
\begin{equation} \label{qdq3}
 Tr(\widehat{U}(x)\widehat{D}(x^{\prime }))=\delta (x,x^{\prime })  \, ,
 \end{equation}
 where $\delta (x,x^{\prime })$ denotes the delta distribution along the diagonal of $X\times X$.
Given an operator $\hat{A}$, the function on $X$ defined as:
\begin{equation}\label{qdq1} 
f_{\hat{A}}(x) = \mathrm{Tr}(\widehat{U}(x)\hat{A}) \, ,
\end{equation}
will be called the symbol of $\hat{A}$ with respect to the dequantizer $\widehat{U}$.  Clearly, because of  (\ref{qdq3}), the operator $\hat{A}$
can be recovered  from its symbol as:
\begin{eqnarray}\label{qdq2}
\hat{A} =\int f_{\hat{A}}(x)\widehat{D}(x) \, \mathrm{d}x \, , 
 \end{eqnarray}%

The star-product of symbols of operators $\hat{A}$ and $\hat{B}$ determined by
 the dequantizer $\widehat{U}(x)$ and the quantizer $\widehat{D}(x)$, 
$ f_{\hat{A}}* f_{\hat{B}} = \mathrm{Tr\,} (U(x)\hat{A}\hat{B})$, 
 is given by the kernel $K(x_{1},x_{2},x_{3})$  \cite{OlgaJPA}:
  \begin{equation}\label{qdq5}
K (x_{1},x_{2},x_{3})= \mathrm{Tr} \left( \widehat{D}(x_{1})\widehat{D}(%
 x_{2})\widehat{U}(x_{3})\right)    \, ,
 \end{equation}%
by means of:
 \begin{equation} \label{qdq4}
 \left( f_{\hat{A}}\ast f_{\hat{B}}\right) (x_{3}) = \int K(x_{1}, x_{2}, x_{3})
 f_{\hat{A}}(x_{1})f_{\hat{B}}(x_{2})\, \mathrm{d}x_{1}\, \mathrm{d}x_{2} \, .
 \end{equation}
  

\subsection{Tomographic picture of quantum states}

We can apply the previous quantizer-dequantizer construction in a different way as we did in the case of Weyl's quantization to get the so called tomographic symbols of the operators  \cite{ManciniPLA96}, \cite{Ibort} and the corresponding associative star-product. 

The tomographic approach uses the dequantizer,
\begin{equation}
\widehat{U}(X,\mu,\nu)=\delta(X\hat{\mathbb{I}}-\mu\hat{q}-\nu\hat{p}) ,\,\,\,\,\,\, X,\mu,\nu \in \mathbb{R}^3,
\end{equation}
and the quantizer,
\begin{equation}
\widehat{D}(X,\mu,\nu)= \frac{1}{2\pi}\exp i(X\hat{\mathbb{I}}-\mu\hat{q}-\nu\hat{p}) \, .
\end{equation}
By using them we define  the symbol-map which associates with any operator  $\hat{A}$ the function
\begin{equation}
\mathcal{W}_{A}(X,\mu,\nu)= \mathrm{Tr}( \hat{A} \, \delta(X\hat{\mathbb{I}}-\mu\hat{q}-\nu\hat{p})) \, ,
\end{equation}
 and 
 \begin{equation}
 \hat{A}=\int \widehat{D}(X,\mu,\nu)\mathcal{W}_{A}(X,\mu,\nu) \, \diff X \diff \mu \diff \nu \, .
 \end{equation}

The kernel of the tomographic star-product reads,
\begin{eqnarray}
&& K\left( X_{1},\mu _{1},\nu _{1},X_{2},\mu _{2},\nu _{2},X_{3},\mu _{3},\nu_{3}\right) = \nonumber \\ && =  \mathrm{Tr} (\widehat{D}(X_{1},\mu_{1},\nu_{1})\widehat{D}(X_{2},\mu_{2},\nu_{2})\widehat{U}(X_{3},\mu_{3},\nu_{3}) )
\end{eqnarray}
and explicitely \cite{OlgaJPA}:
$$
\hspace{-1.0cm} K\left( X_{1},\mu _{1},\nu _{1},X_{2},\mu _{2},\nu _{2},X_{3},\mu _{3},\nu
_{3}\right) = \frac{1}{4\pi ^{2}}\delta \left( \nu _{3}(\mu _{1}+\mu
_{2})-\mu _{3}(\nu _{1}+\nu _{2})\right) \times
$$
\begin{equation}\label{10} 
 \times \exp \left[ i (X_{1}+X_{2})-i\frac{\nu _{1}+\nu _{2}}{\nu _{3}}X_{3}+\frac{i}{2}(\nu _{1}\mu _{2}-\nu _{2}\mu_{1})\right].    
 \end{equation}
This kernel determines the tomographic product of functions $f(X,\mu,\nu)$.


\subsection{The commutative product as a contraction of the Moyal  star-product}\label{sec:moyal}

On the other hand, the pointwise product of functions on the phase-space
\begin{equation}\label{11}
(f_{A}\ast f_{B})(q,p)=f_{A}(q,p)f_{B}(q,p) \, ,  
\end{equation}%
is given by the kernel
\begin{equation}\label{12}
K(q_{1},p_{1},q_{2},p_{2},q_{3},p_{3})=\delta (q_{1}-q_{3})\delta
(q_{2}-q_{3})\delta (p_{1}-p_{3})\delta (p_{2}-p_{3}) \, .
\end{equation}
The kernel of the pointwise product (\ref{12}) is the $\hbar \rightarrow 0$
limit of the Gr\"{o}newald kernel \cite{Gr}:
\begin{equation}\label{L1}
\hspace{-2.2cm} G\left( q_{1},p_{1},q_{2},p_{2},q_{3},p_{3}\right) =\frac{1}{\pi ^{2}\hbar
^{2}}\exp\left( { \frac{2i}{\hbar }\left[
q_{1}p_{2}-q_{2}p_{1}+q_{2}p_{3}-q_{3}p_{2}+q_{3}p_{1}-q_{1}p_{3}\right] } \right).
\end{equation}%
 To see this we use the equivalent form of the Moyal product given by the following formula (see for instance \cite{zachos}) where we have reintroduced the Planck constant,
\begin{equation}
\left( f_{A}\ast f_{B}\right) (q,p)=f_{A}(q,p)\exp \frac{i\hbar }{2}\left(
\overleftarrow{\frac{\partial }{\partial q}}\,\overrightarrow{\frac{\partial
}{\partial p}}-\overleftarrow{\frac{\partial }{\partial p}}\,\overrightarrow{%
\frac{\partial }{\partial q}}\right) f_{B}(q,p)  \, , \label{M1}
\end{equation}%
and where the arrows mean that the
differentiation is applied   to the left factor $f_{A}(q,p)$ or to the
right factor $f_{B}(q,p)$ respectively. 

For $\hbar =0$, Eq. (\ref{M1}) gives the pointwise rule of multiplication described also by
the integral kernel (\ref{12}). In the Moyal formulation of quantum mechanics the
Planck constant $\hbar$ is mathematically interpreted as a
``deformation'' parameter providing a
noncommutative and non-local product on pairs of functions on the phase-space. When the
deformation parameter is zero one has the standard classical mechanics
formalism, with the pointwise commutative and associative product.

It is worthy to note that the explicit procedure to take the
limit $\hbar \rightarrow 0$ needs delicate attention.  The limit is obvious in
the product formula (\ref{M1}), even if it relies on the formal series expansion of the exponential bidifferential operator defining the product, whereas when the
Gr\"{o}newald kernel is considered, the limit in the integral
kernel expression of the product needs an accurate calculation (see below).
The classical limit, $\hbar \to 0$, of the deformed Moyal product was also considered by Thirring \cite{book} by using the properties of Weyl systems.

We will show now that the limit $\hbar \rightarrow 0$ in the Gr\"{o}newald
kernel of the Moyal product provides the kernel of the pointwise product as well.

To do that we will notice first that the geometrical meaning of the term in the exponent of the Gr\"{o}newald kernel Eq. (\ref{L1}), is the area $S$ of the triangle in phase-space with vertices $(q_{1},p_{1}),(q_{2},p_{2})$ and $%
(q_{3},p_{3})$, respectively. Since the area is invariant
with respect to translations, we may consider the triangle
whose vertex $(q_{3},p_{3})$ has been shifted to the origin of the
reference frame under consideration. It means that in the new
system of coordinates
\begin{eqnarray}
\tilde{q}_{1}=q_{1}-q_{3} &,\qquad &\tilde{q}_{2}=q_{2}-q_{3} , \label{L2} \\
\tilde{p}_{1}=p_{1}-p_{3} &,\qquad &\tilde{p}_{2}=p_{2}-p_{3} \, , \nonumber
\end{eqnarray}%
the expression in the exponent takes the form
\begin{equation}
q_{1}p_{2}-q_{2}p_{1}+q_{2}p_{3}-q_{3}p_{2}+q_{3}p_{1}-q_{1}p_{3}=\tilde{q}%
_{1}\tilde{p}_{2}-\tilde{q}_{2}\tilde{p}_{1} \, .  \label{L3}
\end{equation}%
To calculate the $\hbar \rightarrow 0$ limit in the sense of distributions of the Gr\"{o}newald kernel we
use the well-known expression for the Dirac delta function as the limit of the
Gaussian:
\begin{equation}
\lim_{\epsilon \rightarrow 0}\frac{1}{\sqrt{\pi \epsilon }}\exp \left( -\frac{x^{2}%
}{\epsilon }\right) =\delta (x) \, . \label{L4}
\end{equation}%
Here $\epsilon $ is either a real or an imaginary number (this formula maybe 
illustrated, e.g., by the known limit of the free particle propagator at the
zero instant of time:
\begin{equation}
\lim_{t\rightarrow 0}\frac{1}{\sqrt {2\pi it\hbar }}\exp\left[- \frac{
(x-y)^{2}}{2 i t \hbar} \right]=\delta (x-y) \, . ) \label{L5}
\end{equation}%
Then we apply the change of variables in (\ref{L3}) given by the orthogonal
rotation
\begin{eqnarray}
\tilde{q}_{1}=\frac{x_{1}+x_{2}}{\sqrt{2}} &,&\tilde{p}_{2}=\frac{x_{1}-x_{2}%
}{\sqrt{2}} \, ,  \label{L6} \\
\tilde{q}_{2}=\frac{x_{3}+x_{4}}{\sqrt{2}} &,&\tilde{p}_{1}=\frac{x_{3}-x_{4}%
}{\sqrt{2}}\, . \nonumber
\end{eqnarray}%
Thus we get
\begin{equation}
\tilde{q}_{1}\tilde{p}_{2}-\tilde{q}_{2}\tilde{p}_{1}=\frac{1}{2}%
[x_{1}^{2}-x_{2}^{2}-x_{3}^{2}+x_{4}^{2}]\, ,  \label{L7}
\end{equation}%
and for the Gr\"{o}newald kernel as a function of $x_{1},x_{2},x_{3},x_{4}$ we
get the expression:
\begin{equation}
K_{\hbar }(x_{1},x_{2},x_{3},x_{4})=\left[ \frac{\exp
\frac{ix_{1}^{2}}{\hbar }}{\sqrt{i\pi \hbar }}\right] \left[ \frac{\exp
\frac{-ix_{2}^{2}}{\hbar }}{\sqrt{-i\pi \hbar }}\right] \left[ \frac{exp
\frac{-ix_{3}^{2}}{\hbar }}{\sqrt{-i\pi \hbar }}\right]\  \left[ \frac{\exp
\frac{ix_{4}^{2}}{\hbar }}{\sqrt{i\pi \hbar }}\right] . \label{L8}
\end{equation}%
Using the limit formula (\ref{L5}) we get for $\hbar \rightarrow 0$
\begin{equation}
K_{0}(x_{1},x_{2},x_{3},x_{4})=\delta (x_{1})\delta (x_{2})\delta
(x_{3})\delta (x_{4})=\delta (\mathbf{x}) , \label{L9}
\end{equation}%
Here $\mathbf{x}=(x_{1},x_{2},x_{3},x_{4})$.  Finally, using the invariance of the
Dirac delta function under orthogonal transformations:
\begin{equation}
\delta (O\mathbf{x})=\delta (\mathbf{x}) \, , \label{L10}
\end{equation}%
where $O$ is an orthogonal matrix with determinant one, we get, as desired:
\begin{equation}
K_{0}(\tilde{q}_{1},\tilde{p}_{1},\tilde{q}_{2},\tilde{p}_{2})=\delta (%
\tilde{q}_{1})\delta (\tilde{p}_{1})\delta (\tilde{q}_{2})\delta (\tilde{p}%
_{2}).  \label{L11}
\end{equation}%
Substituting back the initial coordinates (\ref{L2}) we get the
pointwise product kernel $K_{0}(q_{1},p_{1},q_{2},p_{2},q_{3},p_{3})=\delta (q_{1}-q_{3})\delta
(q_{2}-q_{3})\delta (p_{1}-p_{3})\delta (p_{2}-p_{3})$.

Turning back to the quantizer-dequantizer formalism discussed above, the quantizer for the Moyal star-product scheme with the Planck constant
reinserted reads,
\begin{equation}
\widehat{D}(q,p)=\frac{1}{\pi \hbar }\exp \left( -\frac{i}{\hbar }(q\hat{p}-p\hat{q}%
)\right) \exp \frac{i}{\hbar }\left( \frac{\hat{q}^{2}+\hat{p}^{2}}{2}-\frac{\hbar }{2}\right) \, ,
\label{L13}
\end{equation}%
and the dequantizer reads,
\begin{equation}
\widehat{U}(q,p)=2\, \pi \hbar \, \widehat{D}(q,p)\, . \label{L14}
\end{equation}%
Since the kernel is not scale invariant with respect to the transform
\begin{equation}
\widehat{U}\rightarrow \lambda \widehat{U},\ \ \ \ \ \ \ \ \widehat{D}\rightarrow
\lambda ^{-1}\widehat{D}  \label{L14bis}
\end{equation}%
while  Eq. (\ref{w8}) is,  the condition for the Gr\"{o}newald
kernel to give in the limit $\hbar =0$ the pointwise product kernel
determines the prefactor $\hbar^{-1}$ for the quantizer $%
\widehat{D}(q,p)$ and the dimensionless prefactor for the dequantizer $\widehat{U}%
(q,p)$.

The previous proof of the $\hbar\to 0$ limit of the Gr\"onewald kernel  is coherent with the mentioned classical proof by Thirring \cite{book}. In fact in \cite{book} it was shown that the Weyl operator $\hat{W}(z)= \exp(z \hat{a}^{\dag} -z^{*}\hat{a}) $ at the limit  $ \hbar\to 0 $ becomes:
\begin{equation}
\hat{W}(z\hbar^{-1/2})\exp (is\hat{q}\hbar^{1/2})\hat{W}(-z\hbar^{-1/2})\to \exp (isq)\hat{\mathbb{I}} \, ,
\end{equation}
where $ q $ is the Weyl symbol of the position operator.
As it was stated in \cite{book} this proposition implies the strong convergence to the identity operator multiplied by a c-number $q$ of the family of operators 
 $ \hat{W}(z\hbar^{-1/2})\hat{q}\hbar^{1/2}\hat{W}^{\dagger} (z\hbar^{-1/2})$. 

To exhibit a situation where more  ``classicality" is  obtained by means of a limiting (contraction) procedure, we consider a system of many particles with different masses (e.g., protons and electrons in an atom). The centre of mass in this system moves according to classical mechanics. The relative motion  obviously has quantum behaviour. Formally it can be explained by introducing the analog of the Wigner-Weyl quantization, not on  phase space $(\mathbf{q},\mathbf{p})$ but on the position-velocity phase space  $ (\mathbf{q},\dot{\mathbf{q}}) $.
 Since $ \dot{q}=p/m $ we have an extra parameter $m$ for any particle.  Because the momentum operator reads as $ \hat{p}=-ih\partial/\partial q $
 the velocity operator contains the ratio $ \hbar/m $.  One has then the possibility to consider the large particle mass limit $ m\to \infty $, instead of the classical limit $ \hbar \to 0 $.   
 
 If we write now the star-product of functions $ f(\mathbf{q},\dot{\mathbf{q}}) $ for bipartite systems $ \mathbf{q}=(q_{1},q_{2}) $ and $\dot{\mathbf{q}}= (\dot{q}_{1},\dot{q}_{2})  $, we get:
 \begin{eqnarray}\label{productstructure}
\hspace{-2cm} \left( f_{1}\ast f_{2}\right) (\mathbf{q},\dot{\mathbf{q}})
= && \nonumber \\ &&  \hspace{-3cm} = \int  f_{1}(\mathbf{q}^{(1)},\dot{\mathbf{q}}^{(1)})   f_{2}(\mathbf{q}^{(2)},\dot{\mathbf{q}}^{(2)}) K(\mathbf{q}^{(1)},\dot{\mathbf{q}}^{(1)},\mathbf{q}^{(2)},\dot{\mathbf{q}}^{(2)},\mathbf{q},\dot{\mathbf{q}}) \, \mathrm{d}\mathbf{q}^{(1)} \mathrm{d}\dot{\mathbf{q}}^{(1)} \mathrm{d}\mathbf{q}^{(2)} \mathrm{d} \dot{\mathbf{q}}^{(2)} \, .
 \end{eqnarray}
 For classical mechanics the kernel above reads:
 \begin{equation*}
\hspace{-2.5cm} K(\mathbf{q}^{(1)},\dot{\mathbf{q}}^{(1)},\mathbf{q}^{(2)},\dot{\mathbf{q}}^{(2)},\mathbf{q}^{(3)},\dot{\mathbf{q}}^{(3)})  = \delta(\mathbf{q}^{(1)}-\mathbf{q}^{(3)})\delta(\mathbf{q}^{(2)}-\mathbf{q}^{(3)})\delta(\dot{\mathbf{q}}^{(1)}-\dot{\mathbf{q}}^{(3)})\delta(\dot{\mathbf{q}}^{(2)}-\dot{\mathbf{q}}^{(3)}) \, .
 \end{equation*}
 For quantum mechanics the kernel reads
 \begin{eqnarray}
    && K(\mathbf{q}^{(1)},\dot{\mathbf{q}}^{(1)},\mathbf{q}^{(2)},\dot{\mathbf{q}}^{(2)},\mathbf{q}^{(3)},\dot{\mathbf{q}}^{(3)}) =  \\
    && \hspace{-2cm} = \prod_{k=1}^{2}  
    \frac{m_{k}^{2}}{ \pi^{2}\hbar^{2}}\exp \frac{2im_{k}}{\hbar}\left( 
  q_{1}^{(k)}\dot{q}_{2}^{(k)}-q_{2}^{(k)}\dot{q}_{1}^{(k)}+  q_{2}^{(k)}\dot{q}_{3}^{(k)}-q_{3}^{(k)}\dot{q}_{2}^{(k)}+  q_{3}^{(k)}\dot{q}_{1}^{(k)}-q_{1}^{(k)}\dot{q}_{3}^{(k)}
  \right) \, , \nonumber 
  \end{eqnarray}
and, as in the situation above, the limit $ \hbar \to 0 $ provides a pointwise kernel. But when we consider the limit $ m_{2} \to \infty $ one has the hybrid kernel form:
   \begin{eqnarray}\label{qc}
      &&K(\mathbf{q}^{(1)},\dot{\mathbf{q}}^{(1)},\mathbf{q}^{(2)},\dot{\mathbf{q}}^{(2)},\mathbf{q}^{(3)},\dot{\mathbf{q}}^{(3)})=  \\ 
      && \hspace{-1.5cm} = \frac{m_{1}^{2}}{ \pi^{2}\hbar^{2}}\exp \frac{2im_{1}}{\hbar}\left( 
    q_{1}^{(1)}\dot{q}_{2}^{(1)}-q_{2}^{(1)}\dot{q}_{1}^{(1)}+  q_{2}^{(1)}\dot{q}_{3}^{(1)}-q_{3}^{(1)}\dot{q}_{2}^{(1)}+  q_{3}^{(1)}\dot{q}_{1}^{(1)}-q_{1}^{(1)}\dot{q}_{3}^{(1)}
    \right) \times \nonumber \\    
&& \times   \delta(\mathbf{q}_{1}^{\;(2)}-\mathbf{q}_{3}^{\;(2)})\delta(\mathbf{q}_{2}^{\;(2)}-\mathbf{ q }_{3}^{\;(2)})\delta(\dot{\mathbf{q}}_{1}^{\;(2)}-\dot{\mathbf{q}}_{3}^{\;(2)})\delta(\dot{\mathbf{q}}_{2}^{\;(2)}-\dot{\mathbf{q}}_{3}^{\;(2)}) \, . \nonumber 
    \end{eqnarray}

 Thus, due to the structure of the kernel we have obtained more ``classicality" in the total system. One degree of freedom is described by a quantum kernel and another one by a classical kernel, as we argued in the Introduction. We can find first order quantum correction to the classical pointwise kernel by expanding  Eq. (\ref{M1}) to first order in $ \hbar $, obtaining a correction term to the pointwise classical kernel of the form:
 \begin{equation*}
\hspace{-2.3cm} \frac{i\hbar}{2m} \left[  \delta' (q_3-q_{1})\delta(\dot{q}_3-\dot{q}_{1})\delta(q_3-q_{2})\delta'(\dot{q}_3-\dot{q}_{2})   \delta (q_3-q_{1})\delta'(\dot{q}_3-\dot{q}_{1})\delta'(q_3-q_{2})\delta(\dot{q}_3-\dot{q}_{2})\right] \,.
\end{equation*} 
\medskip
   
This picture corresponds to the discussion in \cite{Morandi} for the Weyl system described in Lagrangian terms with the Hessian $\partial^{2}\mathcal{L}/\partial \dot{q}_{i}\partial \dot{q}_{j}$ being the ``mass tensor". 
   
    The analogous limit for the tomographic quantum kernel (\ref{10}) follows from the relation of the tomographic symbols to Weyl symbols given by the Radon transform. For a one particle system the classical kernel has the form  given by Eq. (\ref{10}) where the term $ \nu_{1}\mu_{2}-\nu_{2}\mu_{1} $ has been removed. Whereas for two particles with $ m_{1}=1 $ and $ m_{2}\to \infty $ the kernel has a product structure similar to Eq. (\ref{qc}) where the quantum factor is given by  Eq. (\ref{10}) and the classical factor has the same form as  in Eq. (\ref{10}) but without the antisymmetric term in the exponent.


\section{A simple mathematical formulation for the contraction procedure of associative algebras}\label{sec:alternative}


\subsection{The general setting}
Having already discussed in Sect. \ref{sec:weyl} some physical arguments behind the contraction procedure for associative algebras, let us extract the main mathematical ideas 
of this limiting procedure.

Let us consider a bilinear product on a vector space $V$, say $B \colon V \times V \to V$, and a one-parameter family of invertible
linear transformations $T_\lambda \colon V \to V$.  We may define a $\lambda$-dependent bilinear product $B_\lambda$ on $V$ by setting:
 \begin{equation}
B_\lambda (u,v) = T_\lambda^{-1} \left( B(T_\lambda u, T_\lambda v) \right) , \qquad \forall u,v \in V \, .
 \end{equation}

Then, we can construct the trilinear product, called the associator:
 \begin{equation}
A_\lambda (u,v,w) = B_\lambda (B_\lambda (u,v), w) - B_\lambda (u,B_\lambda (v,w)), \qquad \forall u,v,w \in V \, ,
\end{equation}
which defines a $(3,1)$ tensor on $V$.
This tensor vanishes identically (for all $\lambda$) if the product $B$ is associative, that is, the products $B_\lambda$ are associative for all $\lambda$ provided that $B$ is associative.

It may happen, however, that for a critical value of $\lambda = \lambda_c$, $T_c = T_{\lambda_c}$ is not invertible, nevertheless the limit:
\begin{equation}\label{limit}
\lim_{\lambda\to \lambda_c} T_\lambda^{-1} (B(T_\lambda \cdot, T_\lambda \cdot ))  = B_c
\end{equation}
exists.    Therefore in this case, as the transformation $B \to B_c$ does not arise from an invertible linear transformation we cannot automatically conclude that the new product is associative.  If it were, then we would say that $B_cá$ is the associative bilinear product obtained by contraction of $B$ with respect to the family $T_\lambda$.    It is clear that different families $T_\lambda$ could give rise to the same contraction $B_c$ when applied to $B$, for instance, just consider any invertible linear map $S\colon V \to V$ such that $B(Su,Sv) = S(B(u,v))$, then the family $T_\lambda' = T_\lambda \circ S$, will provide the same contraction $B_c$ applied to $B$ than $T_\lambda$. 

 In the paper by Cari\~nena \textit{et al} \cite{CGRM} a condition on $T_\lambda$ was derived, similar to the Nijenhuis condition for Lie algebras, that would guarantee that such limit, if it exists, will provide an associative bilinear product.  

As we are interested in the classical-to-quantum transition our framework as discussed before would be, more specifically, the one of infinite dimensional  $C^*$-algebras, that is, complete normed associative algebras, carrying an antilinear involution $*$ such that $||a^*a|| = || a||^2$ for all elements in the algebra.  The self-adjoint elements in any such algebra, i.e., elements such that $a^*= a$, would be intrepreted as observables of a quantum system.   

In such case the limit in Eq. (\ref{limit}), should be taken with respect to the corresponding topology and in many occasions its computation is  far from trivial.    For this reason it would be more convenient to have  a different approach to describe the contraction mechanism for associative algebras, that is, we may try to describe the contraction limit above directly in terms of structure constants or kernels.   Actually, as we have seen in Sect. \ref{sec:weyl}, even if the computation of the limit is performed at the level of kernels, such computations are more subtle than in the standard, finite dimensional situation.    

Thus we will assume that the associative algebra structure `$\cdot$' is defined on the complex linear space $V$ by the bilinear product $B$, $u\cdot v = B(u,v)$, $u,v \in V$, and we will assume, for the moment, that there is a linear basis $\{ e_i \}$ which allows to write the associative product in terms of structure constants:
\begin{equation}\label{structurec}
e_j \cdot e_k = \sum_l C_{jk}^l e_l \, .
\end{equation}
The associativity of the bilinear product requires that the structure constants $ C^{j}_{mn} $ satisfy
the quadratic equations:
\begin{equation}\label{Q}
\sum_{j} C^{j}_{mn}C^{k}_{jl} = \sum_{j} C^{j}_{nl}C^{k}_{jm} \, .
\end{equation}
We interpret the invertible maps $T_\lambda$ as a change of basis $e_i \mapsto T_\lambda(e_i)$ and, consequently, in the new basis:
\begin{equation}
T_\lambda (e_j) \cdot T_\lambda (e_k) = \sum_l C_{jk}^l(\lambda) T_\lambda (e_l) \, .
\end{equation}
In other words, we define a new product $e_j \cdot_\lambda e_k$ implicitly given by:
\begin{equation}
T_\lambda (e_j \cdot_\lambda e_k ) = T_\lambda (e_j) \cdot T_\lambda (e_k) \, , 
\end{equation} 
with structure constants $C_{jk}^l(\lambda)$, that is, $\sum_l C_{jk}^l(\lambda) e_l = T_\lambda^{-1}(T_\lambda (e_j) \cdot T_\lambda (e_k) )$ and, for the critical value $\lambda_c$, we get the expression equivalent to Eq. (\ref{limit}):
\begin{equation}\label{Climit}
\sum_l C_{jk}^l(\lambda_c) e_l = \lim_{\lambda \to \lambda_c} T_\lambda^{-1}(T_\lambda (e_j) \cdot T_\lambda (e_k) ) \, .
\end{equation}
As it was pointed out before, whether these resulting structure constants satisfy the quadratic condition (\ref{Q}) arising from associativity remains to be checked.

However it is not necessary to introduce a linear basis to obtain an explicit representation of the deformations of an associative product whose limit can be worked out more easily.   As it was shown in the various situations of physical interest in Section \ref{sec:weyl},  a kernel representation provide a convenient setting to perform explicit computations.   Thus, a formal presentation of the contraction method using kernels could be done by assuming that $V$ is a nuclear functional space. Then if the associative product $*$ is described by a continuous bilinear form $B$, then there will exist a kernel $K$ such that
\begin{equation}
(f *g) (x)  = \int  f(x_1) g(x_2) K(x_1, x_2, x) \, \mathrm{d}x_1 \mathrm{d}x_2 \, ,
\end{equation}
and the associativity condition for the kernel function $K(x_1,x_2,x)$ will be expressed as:
\begin{equation}
\int K(x_1,x,x_4)  K(x_2,x_3,x) dx =  \int K(x_1,x_2,x) K(x,x_3,x_4 ) \, \mathrm{d}x \, ,  
\end{equation}
which are the quadratic equations corresponding to Eq. (\ref{Q}).
In the particular instance of a star-product constructed using the quantizer-dequantizer
formalism developed in Section \ref{sec:quantizer}, we get that the kernel is given by  Eq. (\ref{qdq5}).

Hence a family of continuous isomorphisms $T_\lambda$ will induce a family of kernels,
\begin{equation}
B(T_\lambda f, T_\lambda g ) = \int f(x_1) g(x_2)  K_\lambda(x_1, x_2, x) \,  \mathrm{d}x_1 \mathrm{d}x_2 \, ,
\end{equation}
and, finally we will obtain the expression analogous of Eq. (\ref{Climit}):
\begin{equation}
K_c (x_1, x_2, x) = \lim_{\lambda \to 0}  T_\lambda^{-1}(x) K_\lambda(x_1, x_2, x) \, ,
\end{equation}
where the argument $x$ in $T_\lambda$ indicates that it acts on the $x$ variable of the kernel $K_\lambda$.


\subsection{$K$-deformations: A simple example}

In all previous contraction procedures a limiting procedure is required, involving a linear transformation which is not invertible for $\lambda = \lambda_c$. In this section, we show however that it is possible to build alternative associative products by a procedure different from the previous deformation argument.  Consider for any vector $K = \sum_m K^m e_m$ the following product:
\begin{equation}\label{kproduct}
e_j \cdot_K e_k = e_j \cdot K \cdot e_k \, .
\end{equation}
It is clear that if the starting product is associative, the new product $\cdot_K$ is associative too.   By selecting a vector $K = K(\lambda)$ depending on a parameter family $\lambda$, we can define a procedure that in some cases mimics the reduction procedure above.  We will call the associative structure defined by the product $\cdot_K$ defined in Eq. (\ref{kproduct}) a $K$-deformation of the original associative structure.  

We shall elaborate further on this procedure by exhibiting a simple example that illustrates the previous arguments.

On the three dimensional vector space $V = \mathbb{R}^3$ we consider the Lie product:
\begin{equation}
[\mathbf{x}, \mathbf{y}] = \mathbf{x} \wedge \mathbf{y} \, .
\end{equation}
By using the adjoint representation, we associate a $3\times 3$ matrix with any vector.  We denote  by $L_1, L_2, L_3$ the matrices associated to the standard basis of vectors $e_1, e_2, e_3$, i..e,
\begin{equation}
\hspace{-1cm} L_1 = \left( \begin{array}{ccr} 0 & 0 & 0 \\ 0 & 0 & -1 \\ 0 & 1 & 0 \end{array} \right), \quad 
L_2 = \left( \begin{array}{rcc} 0 & 0 & 1 \\ 0 & 0 & 0 \\ -1 & 0 & 0 \end{array} \right), \quad 
L_3 = \left( \begin{array}{crc} 0 & -1 & 0 \\ 1 & 0 & 0 \\ 0 & 0 & 0 \end{array} \right) \, .
\end{equation}
We consider now the standard associative product $A\cdot B$ among matrices in $\mathfrak{gl}(3,\mathbb{R})$  and the deformed $K$-product \cite{K} defined by $A\cdot_K B = A \cdot K \cdot B$, where $K$ is a $3\times 3$ real matrix that doesn't need to be a linear combination of $L_1, L_2, L_3$.
By setting $K$ to be:
\begin{equation}
K =  \left( \begin{array}{ccc} \lambda_1 & \mu_1& \mu_3 \\ \mu_1 & \lambda_2  & \mu_2 \\ \mu_3  & \mu_2  & \lambda_3  \end{array} \right), 
\end{equation}
we get for the commutators $[A,B]_K = A\cdot_K B - B\cdot_K A$:
\begin{eqnarray}
[L_1, L_2]_K &=& \mu_3 L_1 + \mu_2 L _2 + \lambda_3 L_3 , \\ \, 
[L_2, L_3]_K &=& \lambda_1 L_1 + \mu_1 L _2 + \mu_3 L_3, \\ \, 
[L_3, L_1]_K &=& \mu_1 L_1 + \lambda_2 L _2 + \mu_2 L_3 \, .
\end{eqnarray}
All unimodular three dimensional Lie algebras are obtained by properly choosing the parameters appearing in the matrix $K$.   We see that as long as $K$ is invertible, the matrices of these $K$-deformed Lie algebras generate the associative algebra $\mathfrak{gl}(3,\mathbb{R}).$ However when $K$ is not invertible,  we get a ``contraction" from 
 $\mathfrak{gl}(3,\mathbb{R})$ to lower dimensional associative algebras.

On the contrary, if we start with the matrices:
\begin{equation}
\hspace{-1cm} X_1 = \left( \begin{array}{ccc} 0 & 0 & 1 \\ 0 & 0 & 0 \\ 0 & 0 & 0 \end{array} \right), \quad 
X_2 = \left( \begin{array}{ccc} 0 & 1 & 0 \\ 0 & 0 & 0 \\ 0 & 0 & 0 \end{array} \right), \quad 
X_3 = \left( \begin{array}{crc} h & 0 & 0 \\ 0 & 0 & 1 \\ 0 & -1 & 0 \end{array} \right) \, ,
\end{equation}
and
\begin{equation}\label{H}
K =  \left( \begin{array}{ccc} \alpha & \beta & \gamma \\ 0 & \eta  & \rho \\ 0  & \mu  & \nu  \end{array} \right), 
\end{equation}
 the same procedure will provide the remaining family of algebras required to complete the list provided by Bianchi classification of real three-dimensional Lie algebras.
Actually, we get:
\begin{eqnarray}
& & [X_1, X_2]_K = 0 \, , \qquad [X_1, X_3]_K = - \nu X_1+(\mu-h \alpha)X_2  \\
& &  
[X_2, X_3]_K = (\rho+h \alpha)X_1-\nu X_2 \, . 
\end{eqnarray}

Thus we have shown that any contraction of a three-dimensional Lie algebra may be obtained with our alternative procedure of $K$-deformations.


\subsection{$K$-deformations and the Gr\"onewald kernel}\label{sec:Kdeform_Gronewald}
 
 After the discussion of the simple example before, 
we may use the $K$-deformation mechanism to get the new star-product  kernels starting from a given quantizer $\widehat{D}(x)$ and dequantizer $ \widehat{U}(x) $, where $x$ is a point of a measure space $X$ or, for that matter, for any star-product defined by a kernel $K$.   

Then for any given kernel $K$ and function $\kappa(\mathbf{x})$ on the manifold we may define the new $\kappa$-star-product  
 \begin{equation}
 (f_{1}*f_{2})_{\kappa}(x)= (f_{1}*{\kappa}*f_{2})(x)
 \end{equation}
 corresponding to the new kernel
 \begin{equation}\label{deformed_kernel}
\hspace{-1cm} \mathcal{S} (x_{1},x_{2},x_{3})= \int \mathrm{d}z \mathrm{d}y  \, K(x_{1}, 
 z,x_{3})  K(y ,x_{2}, z ) \kappa(y) \, . 
\end{equation}
 Thus, in the quantizer-dequantizer formalism, Eq. (\ref{deformed_kernel}), will become:

 \begin{equation}\label{qdq7}
\hspace{-1cm} \mathcal{S} (x_{1},x_{2},x_{3})= \int \mathrm{d}z \mathrm{d}y  \, \mathrm{Tr} \left(\widehat{D}(x_{1})\widehat{D}( 
 z)\widehat{U}(x_{3})\right) \, 
\mathrm{Tr}\left( \widehat{D}(y )\widehat{D}(x_{2})\widehat{U}(z )\right) {\kappa}(y)  \, .
\end{equation}

As two examples of new star-products for functions on the phase-space, i.e., $x=(q,p)$, where the initial star-product is given by the Gr\"{o}newald kernel Eq. (\ref{L1}), obtained using Eq. (\ref{deformed_kernel}) with functions $ {\kappa}_{1}(q,p)=\delta(q)\delta(p) $ and  $ {\kappa}_{2}(q,p)=\exp (-q^{2}-p^{2}) /\pi$ we will get the new star-products whose kernels will read respectively as:
\begin{equation}
\mathcal{S}_{1}(q_{1},p_{1},q_{2},p_{2},q_{3},p_{3}) = \frac{1}{\pi^{2}} \delta(q_{1}+q_{2}-q_{3})\delta(p_{1}+p_{2}-p_{3})e^{2i(q_{3}p_{1}-p_{3}q_{1})}
\end{equation}
 \begin{equation*}
\hspace{-2.5cm} \mathcal{S}_{2}(q_{1},p_{1},q_{2},p_{2},q_{3},p_{3})= \left( \frac{1}{\pi}\exp [-(q_{1}+q_{2}-q_{3})^{2}-(p_{1}+p_{2}-p_{3})^{2} ]\right)  G(q_{1},p_{1},q_{2},p_{2},q_{3},p_{3}) 
 \end{equation*}
with $G(q_{1},p_{1},q_{2},p_{2},q_{3},p_{3}) $ denoting again Gr\"{o}newald's kernel.

Applying this method $n$ times with different functions $\kappa_{j}(x)$,  $j=1,\dots , n$, one obtains new star-product kernels  which depend  polynomially on the initial one. It is worthy to note that applying the same deformation $ {\kappa}_{1}(q,p)=\delta(q)\delta(p) $ to the above kernel $S_1$ we obtain again the same kernel with an extra factor of normalization, while applying the deformation $ {\kappa}_{2}(q,p)=\exp (-q^{2}-p^{2}) /\pi$ to the above kernel $S_2$ we reconstruct the Gr\"{o}newald's kernel. 

The obtained kernels allow to define new Lie products and new Jordan products.  This construction will be discussed systematically in the following section.


\subsection{Iterated deformations of the associative product}

The construction of the deformed $K$-product  $\cdot_K$ can be iterated by using as a starting associative product $\cdot_K$ itself, thus we would get:
\begin{equation}
e_j \ast_{K_1} e_k = e_j\cdot_K K_1 \cdot _K e_k \, ,
\end{equation}
and so on.   This procedure amounts to define a new product by iteratively using the expression:
\begin{equation}
B_K (u,v) = B(B(u,K), v) = B(u,B(K,v)) \, ,
\end{equation}
the second identity on the r.h.s. of previous equation reflecting the associativity of the original product.  Iterating the construction gives:
\begin{equation}
B_{K_1,K} (u,v) = B_K(B_K(u,K_1), v) = B_K(u,B_K(K_1,v)) \, .
\end{equation}
Clearly, this iteration procedure builds new structure constants in terms of polynomials of the initial
structure constants along with some selected elements $K$, $K_1$, $K_2, \ldots$, in the vector space $V$.
We will change the notation and denote the vectors in the previous sequence by $K^{(0)}, K^{(1)}, \ldots , $ and by $\cdot_{(K_0, K_1, \ldots)}$ the corresponding iterated products.

In a more pictorial way, we could say that the old structure constant 
$a_{mn}^{ r}$ will be contained in the new structure constants, say $s_{mn}^{r}$, in a polynomial form and therefore the original
quadratic condition on the structure constants gives rise to additional
polynomial conditions again satisfied in terms of the original constant structures.


By using the language of vectors associated with matrices (or with functions
in the group algebra setting), our comments would be spelled out in the
following form.

If the structure constants of the original associative product are $a_{jk}^l$, then the structure constants of the product $\cdot_K$
are given by:
\begin{equation}
 C_{mn}^j (K) = \sum_{sp }a_{mp}^s  K^{p} a_{sn}^{j} 
\end{equation}
If we keep iterating the procedure by using the vectors $K^{(1)}, K^{(2)}, \ldots $,  then we will obtain that the structure constants
 for the product $\cdot_{(K,K_1)}$ are given by: 
 \begin{equation}
  S_{mn}^{j}( K^{(1)})= \sum_{sp}C_{mp}^{s} K^{(1)p}C_{sn}^{j} . 
 \end{equation}
 Denoting the structure constants of the iterated product $\cdot_{(K_0, K_1, \ldots, K_r)}$ by $S^{(r)}$ (or, $S^{(r)}(K)$ if we want to emphasize the dependence on the vectors $K^{(s)}$, $s= 1,\ldots, r$) we will have that they are defined recursively by the sequence:
 
\begin{eqnarray*} 
{S^{(1)}} _{mn}^j &=& \sum_{sp}C_{mp}^{s} K^{(1)p}C_{sn}^{j} , \\
{S^{(2)}}_{mn}^j  &=& \sum_{sp} {S^{(1)}} _{mp}^s \, K^{(2)p} \, {S^{(1)}}_{sn}^j \\
 \cdots \cdots && \cdots\cdots \\
{S^{(r)}}_{mn}^j  &=& \sum_{sp} {S^{(r-1)}}_{mp}^s \, K^{(r)p} \, {S^{(r-1)}}_{sn}^j 
\end{eqnarray*}
 
 All the structure constants  ${S^{(r)}}_{mn}^j $ satisfy the associativity equation providing even polynomials equations satisfied by the initial constants $ C_{mn}^{j} $.  As we have seen in the previous subsection, different iterations change the Gr\"{o}newald's kernel to other kernels corresponding to different kinds of quantization of systems in the framework of phase-space functions. Correspondingly the antisymmetric part of the structure constants give polynomial solution of the Jacobi identity, i.e.
\begin{equation}
{L^{(r)}}_{mn}^j(K) = {S^{(r)}}_{mn}^j(K) - {S^{(r)}}_{nm}^j(K)
\end{equation}
satisfies  the Jacobi identity if the initial $ C_{mn}^{j} $ satisfy the associativity equation. 

Now choosing limits  $ K^{(1)}\to K^{(1)}_{0} $, $ K^{(2)}\to K^{(2)}_{0}, \dots $, we get the contracted star-product structure constants and contracted Lie-algebra structure constants.

\subsection{Another example: $SU(2)$ contractions}

Let us consider first the example of the associative algebra associated with the Lie algebra of the  group $U(2)$.  

On the four dimensional vector space of the Lie algebra $\mathfrak{u}(2)$, say $ \mathbb{R}^{4} $, with basis $ e_{0}, e_{j} $, $j = 1,2,3$, we start from
\begin{equation}
 [e_{0},e_{j}]= 0, \qquad [e_{j},e_{k}]=  2 \sum_l \varepsilon_{jkl} e_{l}. 
\end{equation}
 By using the adjoint representation of $\mathfrak{u}(2)$, we obtain the matrix representation $\hat{e}_j$, of the generators $e_j$, i.e.,  $\left(\hat{e}_{j}\right)_{kl}=  2\varepsilon_{(j)kl} $, and we obtain the associative product 
\begin{equation}\label{associative_u2}
\hat{e}_{0}\cdot \hat{e}_{0}= \hat{e}_{0} , \qquad  \hat{e}_{0}\cdot \hat{e}_{j} = \hat{e}_{j} , \qquad \hat{e}_{j}\cdot \hat{e}_{k} = \delta_{jk}\hat{e}_{0} + \sum_l\varepsilon_{jk}^{\phantom{jk}l}\hat{e}_{l}\, .
\end{equation}

Now we consider the linear transformation associated with the matrix:
\begin{equation}\label{Tlambda_u2}
T_\lambda = \left( \begin{array}{cccc}
1 & 0 & 0 & 0 \\
0 &  \lambda& 0 & 0   \\ 
0 & 0 & \lambda & 0 \\ 
0& 0 & 0 & \lambda
\end{array}  \right)  
\end{equation}
and find that we have a new product $ \hat{e}_{\mu}{\cdot}_{\lambda} \hat{e}_{\nu}= T^{-1}_{\lambda}\left( T_\lambda(\hat{e}_{\mu})\cdot T_\lambda (\hat{e}_{\nu})\right) $, $\mu = 0,1,2,3$. 

As it was discussed in detail in Section \ref{sec:alternative} the picture associated with this contraction procedure is as follows.  We start with the binary and bilinear product on a real four-dimensional vector space $V$ given by Eq. (\ref{associative_u2})  with respect to a given basis $\hat{e}_\mu$, $\mu = 0,1,2,3$. then we consider the one-parameter family of linear transformations $T_{\lambda} \colon V\to V$ defined by Eq. (\ref{Tlambda_u2}) and obtain new structure constants by setting $ \hat{e}_\mu{\cdot}_{\lambda} \hat{e}_\nu = T^{-1}_{\lambda}\left( T_{\lambda}(\hat{e}_\mu)\cdot T_\lambda (\hat{e}_\nu)\right) $. For each value of  $ \lambda \neq 0 $, $T_\lambda$ invertible; it turns out that the new product is isomorphic with the old one and therefore satisfies the associativity.  However when $T_\lambda$ is not invertible but somehow the limit  $ \lim_{\lambda\to 0} T^{-1}_{\lambda}\left( T_{\lambda}(\hat{e}_\mu)\cdot T_\lambda (\hat{e}_\nu))\right)  $ still exists it is not guaranteed that the limit structure constants still satisfy the associativity condition.  In the previous situation for $ \lambda=0 $ the transformation is not invertible, but the limit on the right-hand side exists and gives the structure constants of a commutative algebra $\hat{e}_k \cdot_c  \hat{e}_j = 0$, $\hat{e}_0 \cdot_c \hat{e}_\mu = \hat{e}_\mu \cdot_c \hat{e}_0 = \hat{e}_\mu$.

As in Section \ref{sec:alternative} we will consider an alternative way to deform products among matrices and consider these new products as defining contraction procedures.
Following an example studied in \cite{AssociaMoyal} we consider the associative so--called $K -$%
star--product \cite{OlgaJPA, K}, with the matrix multiplication rule $a\circ b=a K b$. In case of $2\times 2-$matrices, by choosing a
Hermitian matrix $K$, we may write 
\begin{equation}
K =\left(  \begin{array}{cc}
k_{11} & k_{12} \\ 
k_{21} & k_{22}
\end{array}\right)  =\sum_{\alpha =0}^{3}s^{\alpha }\sigma _{\alpha }
\end{equation}%
where the components $s^{\alpha },\alpha =0,1,2,3,$ are real, and the $%
\sigma _{\alpha }$ are the Pauli matrices with the identity $\sigma
_{0}$.

The structure constants, with $\alpha =0,1,2,3,$ and $j,m,n=1,2,3$, are:%
\begin{eqnarray*}
C_{00}^{\alpha } &=&s^{\alpha }\, , \qquad C_{0j}^{\alpha }=\left( C_{j0}^{\alpha
}\right) ^{\ast }=\delta _{0}^{\alpha }s^{j}+\delta _{j}^{\alpha
}s^{0}+\delta _{m}^{\alpha }\sum_{n=1}^{3}is^{n}\epsilon _{njm}, \\
C_{jm}^{\alpha } &=&\delta _{0}^{\alpha }\left( s^{0}\delta
_{jm}+\sum_{n=1}^{3}is^{n}\epsilon _{nmj}\right) +\delta _{j}^{\alpha
}s^{m}+\delta _{m}^{\alpha }s^{j}+\delta _{n}^{\alpha }\left( i\
s^{0}\epsilon _{jmn}-\delta _{jm}s^{n}\right) .  \nonumber 
\end{eqnarray*}%

We may consider a contraction product by steps, first $ k_{11}\to 0 $, then $ k_{22}\to 0 $, then $ k_{12}\to 0 $ or $ k_{21}\to 0 $ to finally get a contracted star-product. 

In a similar fashion we may consider the matrices $\eta_{0}=\sigma_0, \eta_{1}=\sqrt{\mu_1 \mu_3}\sigma_1, \eta_{2}=\sqrt{\mu_1 \mu_2}\sigma_2, \eta_{3}=\sqrt{\mu_2 \mu_3}\sigma_3$ whose products depend on parameters $ \mu_{1} $, $ \mu_{2} $, $ \mu_{3} $ as follows:
 \begin{eqnarray*}
  \eta_{1}\eta_{2} &=& - \eta_{2}\eta_{1}=i\mu_{1}\eta_{3}, \quad 
 \eta_{2}\eta_{3}=- \eta_{3}\eta_{2}=i\mu_{2}\eta_{1}, \quad 
\eta_{3}\eta_{1}=- \eta_{1}\eta_{3}=\mu_{3}\eta_{2} \\
\eta_{1}\eta_{1} &=& \mu_{1}\mu_{3}\eta_{0} , \quad 
 \eta_{2}\eta_{2}=\mu_{1}\mu_{2}\eta_{0}, \quad 
 \eta_{3}\eta_{3}=\mu_{2}\mu_{3}\eta_{0}, \quad 
 \eta_{0}\eta_{j}= \eta_{j} , \, \, \, 
  \eta_{0}\eta_{0}= \eta_{0}  .
  \end{eqnarray*}                    
  
 For  each value of the parameters these products are associative and we may consider the limit cases with $ \mu_{1}\to 0 $, then  $ \mu_{2}\to 0 $, and finally  $ \mu_{3}\to 0 $.  Then we will get the product,
 \begin{eqnarray*}
 \mu_{1}\to 0  && \\
 && \eta_{1}\eta_{2}=0 , \quad \eta_{2}\eta_{3}=- \eta_{3}\eta_{2}=i\mu_{2}\eta_{1}, \quad
 \eta_{3}\eta_{1}=- \eta_{1}\eta_{3}=i\mu_{3}\eta_{2} \\
&&  \eta_{1}\eta_{1}=0, \quad 
 \eta_{2}\eta_{2}=0, \quad
 \eta_{3}\eta_{3}=\mu_{2}\mu_{3}\eta_{0} \\
 && \eta_{0}\eta_{j}=\eta_{j},\quad 
   \eta_{0}\eta_{0}= \eta_{0}, \\
 \mu_{2}\to 0  && \\
 &&    \eta_{1}\eta_{2}=0, \quad
     \eta_{2}\eta_{3}=0, \quad 
     \eta_{3}\eta_{1}=\mu_{3}\eta_{2}\\
&&     \eta_{1}\eta_{1}=0, \quad
     \eta_{2}\eta_{2}=0, \quad 
     \eta_{3}\eta_{3}=0 \\
  && \eta_{0}\eta_{j}=\eta_{j} ,\quad
       \eta_{0}\eta_{0}= \eta_{0}, \\
\mu_{3}\to 0  && \\
&&      \eta_{1}\eta_{2}=0,\quad
            \eta_{2}\eta_{3}=0, \quad 
            \eta_{3}\eta_{1}=0\\
  && \eta_{1}\eta_{1}=0, \quad
        \eta_{2}\eta_{2}=0,\quad
        \eta_{3}\eta_{3}=0, \\
   && \eta_{0}\eta_{j}=\eta_{j}, \quad
          \eta_{0}\eta_{0}= \eta_{0} .
\end{eqnarray*}
               
This example shows that the limiting procedure may be considered in a higher dimensional parameter space. This time we start from the observation that for any matrix $ K $, the row-by-column product among matrices, known to be associative, may be  deformed into new associative products by considering $ A{\ast}_k B= A\cdot K \cdot B$ which may be thought of as  a homotopic deformation of the initial one  $ A{\ast}_\lambda B= A\cdot (\mathbb{I}+\lambda ( K-\mathbb{I})) \cdot B$. This class of new associative products is isomorphic with the initial one whenever  $K$ is invertible, when $K$ ceases to be invertible we obtain new associative products not isomorphic to the initial ones.


\section{Conclusions}\label{sec:conclussion}

To conclude we summarise the main results of our work. 
We have considered the quantum-to-classical transition as a contraction procedure of algebraic structures. To this aim,  we have  shown how Wigner-In\"{o}n\"{u}'s contraction of Lie algebras  can be extended to  associative products and we have considered several examples of these situations. 

The contraction procedure introduces Abelian subalgebras or pairwise commuting operators out of non commuting ones. When the contracted algebra may be written as $\mathcal{A}\otimes\mathbb{I}+\mathbb{I}\otimes\mathcal{B}$ with $\mathcal{A}$ Abelian and $\mathcal{B}$ maximally noncommuting, we obtain a system composed by a ``classical" one and a ``quantum" one. 
Starting with the associative product coming from the group algebra of the Heisenberg-Weyl group, we have considered the quantum-to-classical transition, i.e., $ \hbar\to 0 $, as a contraction procedure at the level of ``kernel functions".  Thus the contraction method embodies the phenomena that more ``classicality" can be obtained due to the contraction  procedure very much as more ``Abelianity'' can be obtained as a result of standard contraction of Lie algebras.
We have also discussed the problem of complete and intermediate ``classicality". This procedure is rather involved, as we have shown in the example of Weyl systems.  We have  elaborated on this idea to consider the usual contraction procedure as a way to generate a transition to the classical regime. In particular, by using Weyl systems in the Lagrangian formalism \cite{Dyn},  we have seen that quantum and classical aspects may coexist. In particular, we have discussed how a quantum system containing heavy and light particles, like nuclei and electrons in molecules, may give rise to classical behaviour of heavy particles in the infinite limit of their masses, coupled with quantum behaviour of light particles. However a more complete discussion of these aspects should be considered in forthcoming papers.

Using Ado's theorem, which tells us that any finite dimensional Lie algebra can be realized as a commutator part of a finite-dimensional matrix algebra, Lie algebra contraction can be related to the theory of contractions of associative algebras developed here.  Building on previous work, we have shown that the contraction of the associative algebra gives a contracted associative algebra whose associated Lie algebra is a contracted Lie algebra of the starting one. 

Finally, one of the most important contributions of this work is that by deforming the associative product with the help of elements in the algebra, what we have called $K$-deformations, we are assured that the family of algebras that we define are always associative, therefore it is not necessary to verify the analog of Nijenhuis conditions because they are automatically satisfied. Of course we have not shown that all possible contractions may arise in this way.  However the improvement is extremely important in infinite dimensions because we can avoid dealing with limits in the various topologies we may define on the corresponding spaces of operators.

As for applications to finite level quantum systems, we refer to \cite{Chruscinski:2011yv,Chrubis} where the contraction procedure is related to the semigroup associated with the Kossakowski-Lindblad Markovian dynamics.  This and other applications involving contractions of groupoid algebras will be discussed elsewhere.

 

\ack
A.I. was partially supported by the Community of Madrid project QUITEMAD+, S2013/ICE-2801, and MINECO grant MTM2014-54692-P.   V.M. would like to thank the warm hospitality of the Dipartimento di Fisica dell'Universita di Napoli were this work was started.


\end{document}